\documentclass[12pt]{article}
\newcommand{\sect}[1]{\setcounter{equation}{0}\section{#1}}
\newcommand{\eq}{\begin{equation}}
\newcommand{\eqa}{\begin{eqnarray}}
\newcommand{\en}{\end{equation}}
\newcommand{\ena}{\end{eqnarray}}
\newcommand{\enn}{\nonumber \end{equation}}

\def\sk{\vskip .4cm}
\def\noi{\noindent}

\def\al{\alpha}

\let \si\sigma
\let \part\partial

\def\unmezzo{{1 \over 2}}

\def\de{\delta}

\def\part{\partial}

\def\sk{\vskip .4cm}

\def\noi{\noindent}

\def\X0{X^0}

\def\al{\alpha}

\def\unmezzo{{1 \over 2}}

\def\de{\delta}

\def\La{\Lambda}

\def\dmup{\part^{\mu}}

\def\viel#1#2{e^{#1}_{~~{#2}}}

\def\square{{\,\lower0.9pt\vbox{\hrule \hbox{\vrule height 0.2 cm
\hskip 0.2 cm \vrule height 0.2 cm}\hrule}\,}}

\def\Lcal{{\cal L}}

\def\dmu{{\part_\mu}}
\def\dnu{{\part_\nu}}
\def\dnup{{\part^\nu}}

\def\viel#1#2{e_{#1}^{~#2}}
\def\Xtilde{\tilde X}
\def\xhat{\hat x}
\def\ds{\stackrel{\star}{,}}

\def\Phi{\phi}
\def\rf{{\rm f}}

\begin{document}

\begin{titlepage}
\rightline{DISTA-UPO/08}
%\rightline{hep-th/9509031}
\rightline{March 2008}
\vskip 2em
\begin{center}{\bf DYNAMICAL NONCOMMUTATIVITY\\[.5em] AND NOETHER THEOREM
IN TWISTED $\Phi^{\star 4}$ THEORY}
\\[3em]
{\bf Paolo Aschieri}${}^{1,2}$, {\bf Leonardo Castellani}${}^{2}$,
{\bf Marija Dimitrijevi\' c}${}^{2,3}$ \\ [3em] {\sl ${}^{1}$
Centro ``Enrico Fermi", Compendio Viminale, 00184 Roma, Italy \\
[1em] ${}^{2}$ Dipartimento di Scienze e Tecnologie avanzate and
\\ INFN Gruppo collegato di Alessandria,\\Universit\`a del Piemonte Orientale,\\ Via Bellini 25/G 15100
Alessandria, Italy}\\ [.5 em] {\sl ${}^{3}$Faculty of Physics,
University of Belgrade, \\ P. O. Box 368, 11001 Belgrade, Serbia}
\\[3em]

\end{center}

\begin{abstract}
A $\star$-product is defined via a set of commuting vector fields $X_a
= \viel{a}{\mu} (x) \part_\mu$, and used in a $\Phi^{\star 4}$ theory
coupled to the $\viel{a}{\mu} (x)$ fields. The $\star$-product
is dynamical, and the vacuum solution $\Phi =0$,
$\viel{a}{\mu}=\de_a^\mu$ reproduces the usual Moyal product. The
action is invariant under rigid translations and Lorentz
rotations, and the conserved
energy-momentum and angular momentum tensors
are explicitly derived.

\end{abstract}

\vskip 6cm \noi \hrule \vskip.2cm \noi {\small
leonardo.castellani@mfn.unipmn.it\\ aschieri, dimitrij@to.infn.it}

\end{titlepage}

\newpage
\setcounter{page}{1}

\sect{Introduction}

Noncommutative coordinates are a recurrent theme in mathematical
physics. Early considerations on quantum phase space geometry can
be found in \cite{Dirac}, and the idea of noncommuting spacetime
coordinates goes back to Heisenberg who suggested (in a letter to
Peierls \cite{Heis}) that uncertainty relations between spacetime
coordinates could resolve the UV divergences arising in quantum
field theories. This motivation still holds today, in particular
for nonrenormalizable  field theories of gravity where
 finiteness is the only option for consistency.

The issue was explored initially by Snyder in \cite{Snyder}, and
since then noncommutative geometry has found applications in many
branches of physics, in particular in the last two decades. Some
comprehensive reviews can be found in references \cite{Connes},
\cite{Landi}, \cite{madorebook},
  \cite{Castellani1}, \cite{revDN},\cite{revSz1}, \cite{revSz2}. As an
important example, the development of the noncommutative
differential geometry on quantum groups (continuous deformations
of Lie groups) and more generally on Hopf algebras, has led to
interesting generalizations of gauge and gravity theories, whose
symmetries are deformations of the corresponding classical
symmetries (see for ex.  \cite{AC},\cite{Castellaniqgauge},
\cite{Castellaniqgrav},\cite{grav1},
\cite{grav2},\cite{twgauge},\cite{vasmoz}).

On the other hand, string theories have been pointing towards a
non-commuting scenario already in the 1980's \cite{Witten1}. Later
Yang-Mills theories on noncommutative spaces have emerged in the
context of M-theory compactified on a torus with a constant
background 3-form field, or as the low-energy limit of open
strings in a background B-field describing the fluctuations of the
D-brane worldvolume \cite{SW}.

Noncommutative spacetime is described in terms of coordinates
$\xhat^\mu$,
 \eq [\xhat^\mu,\xhat^\nu]= i
\theta^{\mu\nu} \label{xcomm}
\en
\noi where $\theta^{\mu\nu}$ is an antisymmetric tensor, usually chosen
to be constant (corresponding to constant background fields in string theory).

The algebra of functions of these noncommuting coordinates
can be represented by the algebra of functions on {\it ordinary}
spacetime, equipped with a noncommutative $\star$-product. For constant
 $\theta^{\mu\nu}$ it is known as the Groenewold-Moyal
product \cite{groenewold}, \cite{MW}:
 \eq\label{deformationquant}
 f(x) \star g(x) \equiv  e^{\Delta}
(f,g),~~~\Delta (f,g) \equiv {i \over 2} \theta^{\mu\nu} (\dmu f
)(\dnu g).
\en

\noi and indeed reproduces (\ref{xcomm}) as $x^\mu \star x^\nu - x^\nu \star x^\mu = i
\theta^{\mu\nu}$. This $\star$-product is associative and
noncommutative, and was first introduced to represent on the
classical phase space the product of quantum operators.

Field theories on noncommutative spacetime can then be obtained by
replacing the usual product between fields with the
$\star$-product. Because of the non-polynomial character of the
$\star$-product the resulting field theories are non-local.\footnote{The realization
(\ref{deformationquant}) of the $\star$-product $f\star g$
holds for a limited class of functions (e.g.
polynomials, or analytic and rapidly decreasing functions).
For a richer class of functions, e.g. smooth and rapidly decreasing (Schwarz test functions), an integral representation of the $\star$-product is needed. One such representation is
$f\star g\,(x)=(2\pi)^{-2D}\int\!\!\int
f(x+{1\over 2}\theta u)g(x+s)e^{ius}{\rm d}^Du\,{\rm d}^Ds\,$  \cite{nonlocalstar}
and explicitly encodes the nonlocality of the $\star$-product.}
Thus deformations of scalar,
Yang-Mills, gravity theories have been
considered, and (at least in the first two cases) their quantum
behaviour is under active investigation, see for ex.
\cite{NCrenorm,GSW} and references therein.

The $\star$-deformation
usually leads to $\star$-deformed invariances of the non-local
actions: for example $U(n)$ $\star$-Yang-Mills theory is invariant
under $\star$-gauge transformations on the fields $\de_\epsilon
A_\mu = \dmu \epsilon -i(A_\mu \star \epsilon - \epsilon \star
A_\mu)$.
Spacetime symmetries are likewise deformed: while the non-local
actions are invariant under rigid translations (so that a
conserved energy momentum tensor can be found via the usual
Noether theorem), Lorentz symmetry is typically broken, since the
constant antisymmetric tensor $\theta^{\mu\nu}$ cannot be a
Lorentz invariant tensor in $D > 2$, cf.
\cite{Micu},\cite{Gerhold},\cite{Pengpan},\cite{AbouZeid},\cite{Gonera},
and \cite{GLRV}.
However the $\star$-deformed action is invariant under a deformed
Lorentz symmetry, acting on $\star$-products of fields with a
deformed Leibniz rule \cite{twistPoinc,Gonera}. In this case the
usual Noether theorem for global Lorentz rotations does not apply, and no conserved charge has
been found so far.
\sk
In the present paper we propose a way to restore exact
(undeformed) Lorentz symmetry in a $\star$-deformed interacting
scalar theory. The key ingredient is a generalized Moyal product,
defined via a set of commuting vector fields $X_a = \viel{a}{\mu}(x)
\dmu$ as given in  eq. (\ref{twist}). This product corresponds
to the twist ${\cal F} = \exp [-{i \over 2} \theta^{ab} X_a \otimes X_b]$.
It gives rise to a twisted scalar field theory where $\viel{a}{\mu}$,
and hence the $\star$-product itself, becomes dynamical. The condition $[X_a,X_b]=0$
implies constraints on $\viel{a}{\mu}$, that can be solved off-shell in terms of
$D$ scalar fields $\phi^a$. Thus the dynamical $\star$-product is well defined
off-shell.\footnote{ A $\phi^{\star 4}$ action with spacetime dependent (but {\it nondynamical})
noncommutativity has been considered in \cite{Ruiz}.} Field theories on
noncommutative spaces  (\ref{xcomm}) with  nonconstant
$\theta^{\mu\nu}(x)$ have been considered for example in ref.s \cite{JSW,Sitarz,grav2,GSW}.

In Section 2 we determine the $\viel{a}{\mu}$ fields in terms of the $\phi^a$, and
specify the action of the $\Phi^{\star 4}$ theory coupled to the $\Phi^a$ fields.
Section 3 contains the
variations of the Lagrangian and the resulting field equations.
In Section 4 Noether theorem is applied to derive conserved
energy-momentum and angular momentum tensors. Section 5 contains
some final considerations. Useful formulas are collected in the
Appendix.

\sect{Dynamically twisted $\Phi^{\star 4}$ theory }

\subsection{Generalized Moyal product}

One of the most studied examples of noncommutative (deformed)
spaces is the canonically deformed space or the $\theta$-deformed
space, see for ex. \cite{revSz1}.  The deformation is contained in
the associative and noncommutative Groenewold-Moyal
$\star$-product given by
\begin{equation}
f\star g = \mu \big{\{} e^{\frac{i}{2}\theta^{\rho\sigma}\partial_\rho \otimes\partial_\sigma}
f\otimes g \big{\}} , \label{MWstar}
\end{equation}
where the map $\mu$  is the usual pointwise
multiplication: $\mu (f \otimes g)= fg$. This product can be
generalized as
\begin{eqnarray}
f\star g &=& \mu \big{\{} {\cal F}^{-1} f\otimes g \big{\}}
\nonumber\\ &=& \mu \big{\{} e^{\frac{i}{2}\theta^{ab}X_a \otimes
X_b} f\otimes g \big{\}}  \nonumber\\ &=& e^\Delta (f,g)
,\label{twist}
\end{eqnarray}
where
 $\theta^{ab}$ is a constant antisymmetric matrix, and the
 bilinear operator $\Delta$ is defined by (\ref{twist}) to act on
 a couple of functions as
 \eq
 \Delta (f,g) \equiv {i \over 2} \theta^{ab} (X_a
f) (X_b g) ~.
 \en
 \noi (cf. Appendix). The twist ${\cal F}$ is given by
\begin{equation}
{\cal F} = e^{-\frac{i}{2}\theta^{ab}X_a \otimes X_b} \label{twistdef}
\end{equation}
and $X_a = \viel{a}{\mu} (x) \part_\mu $ are $D$ commuting vector fields, the index $a$ being just a label for the vector fields. The coordinates $x^\mu$ span a  $D$-dimensional Minkowski space with metric $\eta_{\mu\nu}$. In the commutative limit  ($\theta^{ab}\to 0$) the product (\ref{twist}) reduces to the usual pointwise multiplication. The requirement that  the vector fields $X_a$ commute ensures the associativity of (\ref{twist}). From
\begin{equation}
[X_a, X_b ] =0 \label{Xcom}
\end{equation}
we obtain the condition
\begin{equation}
\viel{[a}{\nu} \dnu \viel{b]}{\mu} =0 .\label{vielcon1}
\end{equation}
Supposing that the square matrix $\viel{a}{\mu}$ has an inverse
$\viel{\mu}{a}$ everywhere (so that the $X_a$ are linearly independent),
the condition becomes $\part_{[\mu}
\viel{\nu]}{a}=0$ and it is solved by
\eq
\viel{\nu}{a} (x) = \partial_\nu \Phi^a
(x) . \label{vielcon2}
\en
In this way the $\star$-product (\ref{twist}) is determined by $D$
scalar fields subject to the condition that $\dmu \Phi^a$ is
everywhere invertible. Since $X_a \Phi^b = \de^b_a $, the fields
$\Phi^b$ can be seen as new coordinates along the $X_a$
directions. In particular, from (\ref{twist}) we find
\begin{equation}
\lbrack x^\mu \ds x^\nu \rbrack \equiv x^\mu \star x^\nu -
  x^\nu \star x^\mu = i\theta^{ab}\viel{a}{\mu}\viel{b}{\nu} .\label{xcom}
\end{equation}
Noncommutativity is here given by the space-time dependent (possibly degenerate)
antisymmetric tensor
\eq
 \Theta^{\mu\nu}(x)\equiv\theta^{ab}\viel{a}{\mu}(x)\viel{b}{\nu}(x)~.
 \label{theta}
  \en
With this particular form of $x$-dependent noncommutativity
parameter, originating from the twist (\ref{twistdef}), we have at
our disposal the powerful twist machinery that allows to construct
the differential calculus and geometry \cite{grav2} relevant for
the $\star$-product (\ref{twist}). The dimensionful parameters
$\theta^{ab}$ can be considered fundamental constants (for example
related to Planck length). Note also that $\Theta^{\mu\nu}$
transforms as an antisymmetric tensor under usual Lorentz
transformations $x^\mu\rightarrow \La^\mu_{~\nu} x^\nu$, since
 $\viel{a}{\mu}(x)$ transforms as a vector. The relation
$\lbrack x^\mu \ds x^\nu \rbrack=i
\Theta^{\mu\nu}(x)$ is therefore covariant under usual Lorentz transformations.
 The
$\star$-product is invariant under Lorentz transformations,
as it follows most easily from the
Lorentz invariance of the vector fields
$X_a=\viel{a}{\mu}(x)\partial_\mu$.

Due to (\ref{Xcom}) the action of
$X_a$ satisfies the Leibniz rule:
\begin{equation}
X_a \big( f\star g \big) = (X_a f)\star g + f\star (X_a g) \label{LruleX}
\end{equation}
\noi whereas  a {\sl deformed} Leibniz rule holds for the usual partial derivatives
 $\partial_\mu = \frac{\partial}{\partial x^\mu}$ \cite{grav2} .

\subsection{Action}

We use the $\star$-product (\ref{twist}) to define an action for  $\Phi^{\star 4}$ theory coupled to
$\Phi^c$ :
\begin{eqnarray}
S[\Phi , \Phi^a] &=& \int \Big( \unmezzo \dmu \Phi \star \dmup
\Phi - {m^2 \over 2} \Phi \star \Phi - {\lambda \over 4 !} \Phi
\star \Phi \star \Phi \star \Phi  \nonumber\\ && \quad \quad +
\unmezzo \dmu \Phi_c \star \dmup \Phi^c \Big)  ~ {\mbox{d}}^D x ~
. \label{action1}
\end{eqnarray}
\noi Note however that the above integral is not cyclic: even with
suitable boundary conditions at infinity

\begin{equation}
\int (f\star g) ~ {\mbox{d}}^D x \not= \int (g\star f)~ {\mbox{d}}^D x
\end{equation}

\noi  since $f \star g =  g \star f + X_a (G^a)$ (see formula
(\ref{app2}) in the Appendix where $G^a$ is given explicitly) and
$X_a(G^a)$ is {\sl not} a total derivative. A cyclic integral can
easily be defined by using the measure $e\,{\mbox{d}}^D x $ where
$e=\det (\viel{\mu}{a})$. Indeed $e X_a(G^a) = \dmu (e
\viel{a}{\mu} G^a)$ for any $G^a$, so that up to boundary terms:

\eq \int (f\star g) \,e\, {\mbox{d}}^D x=\int fg \,e\,{\mbox{d}}^D
x = \int (g\star f)\,e\, {\mbox{d}}^D x  ~.\label{fsgfggsf}
\en

 The action (\ref{action1}) can then be
rewritten by means of a cyclic integral:

\begin{eqnarray}
S[\Phi , \Phi^a] &=& \int \Big[ \Big( \unmezzo \dmu \Phi \star
\dmup \Phi - {m^2 \over 2} \Phi \star \Phi - {\lambda \over 4 !}
\Phi \star \Phi \star \Phi \star \Phi  \nonumber\\ && \quad \quad
+ \unmezzo \dmu \Phi_c \star \dmup \Phi^c \Big)\star e^{-1} \Big]
e\,{\mbox{d}}^D x ~.\label{action}
\end{eqnarray}
\noi

 \noi Equation (\ref{fsgfggsf})
%(or the identity (\ref{app1}) in the Appendix)
allows to remove the $\star$-product in
$\star~ e^{-1}$ and proves the equality of (\ref{action1}) and
(\ref{action}).
% up to boundary terms.

Unlike the ordinary Moyal case, the absence of the quartic
potential $\Phi^{\star 4}$  does {\it not} correspond to a free scalar theory, since
only one $\star$-product can be removed in the remaining terms of
(\ref{action}). Then $\theta$-dependent terms involve higher-order
couplings between $\Phi$ and $\Phi^c$ fields.

 World-index contractions being defined with the
Minkowski metric $\eta^{\mu\nu}$, the action is invariant under
global Lorentz transformations.

\sect{Variation of the Lagrangian and field equations}

\subsection{$\Phi$ variation}

We now derive the equations of motion for the fields $\Phi$ and $\Phi^a$. To vary the action (\ref{action}) with respect to the field $\Phi$ we use the usual Leibniz rule, for example
\begin{equation}
\delta_\Phi \Big( -\frac{m^2}{2}\int \big( \Phi\star\Phi\star e^{-1} \big) e\, {\mbox{d}}^D x\Big) = -\frac{m^2}{2}\int \big( (\delta\Phi\star\Phi + \Phi\star\delta\Phi)\star e^{-1} \big) e\, {\mbox{d}}^D x .\nonumber
\end{equation}
The varied terms are grouped in the following way
\begin{eqnarray}
\delta_\Phi S &=& \delta_\Phi \int \big( {\cal L}_\star \star e^{-1} \big) e\, {\mbox{d}}^D x \nonumber\\
&=& \int \big( \de \Phi ~ E_{\Phi}  +
\dmu K^{\mu} \big) ~ {\mbox{d}}^D x , \label{phivariation}
\end{eqnarray}
with ${\cal L}_\star$ defined as
\begin{equation}
{\cal L}_\star = \unmezzo \dmu
\Phi \star \dmup \Phi - {m^2 \over 2} \Phi \star \Phi - {\lambda \over 4
!} \Phi \star \Phi \star \Phi \star \Phi
+ \unmezzo \dmu \Phi_c \star \dmup \Phi^c .\label{Lstar}
\end{equation}

The equations of motion for the field $\Phi$ are:
\begin{equation}
E_{\Phi} = \frac{1}{2}\partial_\mu\Big(e\{\partial^\mu\Phi \ds e^{-1}\} \Big)
+ \frac{m^2}{2}e\{ \Phi \ds e^{-1}\} + \frac{\lambda}{4!} e\{\Phi\star\Phi \ds \{\Phi \ds e^{-1}\} \} =0 . \label{phiequation}
\end{equation}
In the commutative limit $\theta^{ab}\rightarrow 0$ this equation
reduces to the usual field equation for the $\Phi^4$ theory
\begin{equation}
\square \Phi + m^2\Phi +\frac{\lambda}{3!} \Phi^3 =0 .\label{phieomcl}
\end{equation}

The current $K^{\mu}$ is given by
\begin{eqnarray}
K^{\mu} &=& \frac{e}{2}\de\Phi \{\partial^\mu\Phi \ds e^{-1}\} \nonumber\\
&& + e \viel{a}{\mu}\Big[ T(\Delta) \big( \dnu \de \Phi, \unmezzo \Xtilde^a
\{\dnup \Phi \ds e^{-1} \} \big) \nonumber \\
&& \quad\quad- \frac{m^2}{2}T(\Delta)\big( \de \Phi, \Xtilde^a \{\Phi \ds e^{-1} \}
\big) \nonumber\\
&&\quad\quad - \frac{\lambda}{4!}T(\Delta)\Big( \de \Phi, \Xtilde^a \big(\{\Phi^{\star2} \ds \{\Phi \ds e^{-1} \} \} \big)\Big) \nonumber \\
&&\quad\quad + S(\Delta) \big( \dnu\Phi, \Xtilde^a ( \dnup \de \Phi \star e^{-1})\big)
\nonumber \\
&&\quad\quad - m^2 S(\Delta) \big(\Phi , \Xtilde^a (\de \Phi \star e^{-1}) \big) +
\frac{\lambda}{12} S(\Delta) \big(\Phi , \Xtilde^a (\de \Phi \star \Phi^{\star2} \star e^{-1})\big) \nonumber \\
&& \quad\quad - \frac{\lambda}{12}S(\Delta) \big(\Phi \star \Phi, \Xtilde^a (
\de \Phi \star \Phi \star e^{-1} )\big) \nonumber \\
&&\quad\quad - \frac{\lambda}{12}S(\Delta) \big( \Phi \star \Phi \star \Phi , \Xtilde^a (
\de \Phi \star e^{-1} )\big) \Big]  \label{phicurrent}
\end{eqnarray}
where $T(\Delta)$ and $S(\Delta)$ are operators defined in terms of $\Delta$:
\eq
T(\Delta) \equiv {\exp( \Delta) -1 \over \Delta} , ~~~~S(\Delta)
\equiv {\sinh \Delta \over \Delta} \label{TSDelta}
\en
and $\Xtilde^a \equiv {i\over 2} \theta^{ab} X_b$. Useful
identities for the derivation of $E_\Phi$ and $K^\mu$ are given in
the Appendix.

\subsection{$\Phi^c$ variation}

The variation of the action (\ref{action}) with respect to the
field $\Phi^a$ has to be carried out carefully, since the field
$\Phi^a$ appears in the $\star$-product as well. One useful rule
is (see the Appendix):
\eq \de_{\Phi^c} (f \star g) = - (\de \Phi^c
X_c f) \star g - f \star  (\de \Phi^c X_c g) + \de \Phi^c X_c ( f
\star g) \label{varstar}
\en
\noi where the functions $f$ and $g$ do not depend on $\Phi^c$. We
group the terms in the following way
\begin{eqnarray}
\delta_{\Phi^c} S &=& \delta_{\Phi^c} \int \big( {\cal L}_\star \star e^{-1} \big) e\, {\mbox{d}}^D x \nonumber\\
&=& \int \big( - \de \Phi^c
(X_c \Phi) E_{\Phi} + \de \Phi^c E_{\Phi^c} + \dmu J^{\mu} \big) ~ {\mbox{d}}^D x . \label{phicvariation}
\end{eqnarray}
Therefore the equations of motion for the field $\Phi^c$ read:
\begin{equation}
- (X_c \Phi) ~E_\Phi + E_{\Phi^c} =0
\end{equation}
with
\begin{eqnarray}
E_{\Phi^c}  &=& - \frac{1}{2}\dmu \big( e \{\dmup\Phi_c \ds
e^{-1}\} \big) + \frac{1}{2}(X_c\Phi)\dmu \big( e \{\dmup\Phi \ds
e^{-1}\} \big)\nonumber \\
&& +\frac{e}{2}(X_c\partial_\mu\Phi^a)\{\dmup\Phi^a \ds
e^{-1}\} + \frac{e}{2}(X_c\partial_\mu\Phi)\{\dmup\Phi \ds
e^{-1}\} -X_c {\cal L}_\star  .\label{phicequation}
\end{eqnarray}
When $\Phi$ is on shell (i.e. $E_\Phi =0$), the $\Phi^c$ field equations become simply
\eq
E_{\Phi^c} =0 \label{clphicequation}
\en
and reduce in the commutative limit  to
\begin{equation}
\square \Phi^c =0. \label{phiceomcl}
\end{equation}
Note that the field equations
(\ref{phiequation}),(\ref{phicequation}) are satisfied by the
vacuum solution $\Phi = 0$, $\viel{\mu}{a} \equiv \dmu \Phi^a =
\de^a_\mu$ (corresponding to the usual Moyal product). The field
$\Phi$ acts as a source for the noncommutativity field $\Phi^c$.

The current $J^\mu$ reads
\begin{eqnarray}
J^\mu &=& K ^\mu (\de \Phi \rightarrow - \de \Phi^c X_c \Phi) \nonumber \\
&& + {e \over 2} \de \Phi_c \{ \dmup \Phi^c \ds e^{-1} \}
+ {e \over 2} \de \Phi^c (X_c \Phi) \{ \dmup \Phi \ds e^{-1} \} \nonumber \\
&& + e  \viel{a}{\mu} \big(\de \Phi^a (\Lcal_\star \star e^{-1}) -
\Lcal_\star \star (\de \Phi^a e^{-1})\big) \nonumber \\
&& +  e  \viel{a}{\mu} \Big[ T(\Delta) \big( X_c \Lcal_\star, \Xtilde^a (\de
\Phi^c e^{-1})\big) \nonumber \\
&& \quad\quad + T(\Delta) \Big(\part_{\si}(\de \Phi^c \viel{c}{\rho})
\part_{\rho} \Phi, \unmezzo \Xtilde^a \big( \{ \part^{\si} \Phi \ds e^{-1} \} \big)
\Big) \nonumber \\
&& \quad\quad + T(\Delta) \Big( \part_{\si}(\de \Phi^c \viel{c}{\rho})
\part_{\rho} \Phi_d, \unmezzo \Xtilde^a \big( \{ \part^{\si} \Phi^d \ds e^{-1} \} \big)
\Big) \nonumber \\
&& \quad\quad+ S(\Delta) \Big( \part_{\si} \Phi, \Xtilde^a \big(
(\part^{\si}(\de \Phi^c \viel{c}{\rho}) \part_{\rho} \Phi) \star e^{-1} \big) \Big)
\nonumber \\
&& \quad\quad + S(\Delta) \Big( \part_{\si} \Phi_d, \Xtilde^a \big(
(\part^{\si}(\de \Phi^c \viel{c}{\rho}) \part_{\rho} \Phi^d) \star e^{-1} \big) \Big) \Big] .
\label{phiccurrent}
\end{eqnarray}

\sect{Symmetries and conserved currents}

Under a functional variation of the fields and a coordinate
change: \eqa & &\Phi ' (x) = \Phi (x) + \de \Phi (x)
\label{varphi}\\
& &\Phi^{'c} (x) = \Phi^c (x) + \de \Phi^c (x)
\label{varphic}\\
& & {x'}^\mu = x^\mu + \epsilon^\mu \label{varx}
\ena
the variation of the action, to first order in $\de
\Phi(x)$, $\de \Phi^c(x)$ and $\epsilon^\mu (x)$, is:
\eq
\de S=\int \Big(\de_{\Phi} [(\Lcal_\star \star  e^{-1}) e] + \de_{\Phi^c}
[(\Lcal_\star \star  e^{-1}) e]+ \epsilon^\mu \dmu  [(\Lcal_\star \star  e^{-1}) e]
+ (\Lcal_\star \star  e^{-1}) e ~
\dmu \epsilon^\mu \Big)~ {\mbox{d}}^D x\label{varS} \en

\noi where we have used ${\mbox{d}}^D x' = [1 + \dmu \epsilon^\mu
+ O(\epsilon^2)] {\mbox{d}}^D x$.

On shell, and integrated on an arbitrary manifold $M$ (so that
the total derivative terms do not disappear), this
variation takes the form:

\eq \de S = \int_M \dmu [K^\mu + J^\mu +  \epsilon^\mu
(\Lcal_\star \star e^{-1}) e ]~{\mbox{d}}^D x .
\label{varSonshell}\en

\subsection{Energy-momentum tensor}

The action (\ref{action}) is invariant under global translations,
i.e. under the transformations:

\eq
\de \Phi = - \epsilon^\nu \dnu \Phi,~~~\de \Phi^c = - \epsilon^\nu \dnu
\Phi^c,~~~\epsilon^\nu = constant \>.
\en
Substituting these variations into (\ref{varSonshell}) leads to
 \eq
  0=\de S= \int_M \epsilon^\nu \dmu T^{\mu}_{~\nu} ~{\mbox{d}}^D x .
   \en
 \noi where
 \begin{eqnarray}
T^{\mu}_{~\nu} &=& -{e \over 2} (\dnu \Phi) \{ \dmup \Phi \ds
e^{-1} \} -{e \over 2} (\dnu \Phi_c) \{ \dmup \Phi^c  \ds e^{-1} \} \nonumber \\
& & + e \viel{a}{\mu} \Big( \Lcal_\star \star (e^{-1} \dnu \Phi^a )
- T(\Delta) (X_c \Lcal_\star, \Xtilde^a ( e^{-1}\dnu \Phi^c)) \Big) \label{Tmunu}
\end{eqnarray}
\noi is the conserved energy-momentum tensor. This tensor is not symmetric: only
in the commutative limit
it reduces to the canonical (and symmetric) energy-momentum tensor
of the decoupled $\Phi$ and $\Phi^c$ fields.

{}From (\ref{Tmunu}) the divergence $\dmu T^{\mu}_{~\nu}$ can be
explicitly computed and shown to vanish on shell.

\subsection{Angular momentum tensor}

The action  (\ref{action}) is invariant under global Lorentz
rotations, i.e. under the transformations:
\eq
\de \Phi = - \epsilon^\nu \dnu \Phi= - \epsilon^{\nu \rho} x_\rho
\part_\nu \Phi,~~~~\de \Phi^c = - \epsilon^\nu \dnu \Phi^c=
- \epsilon^{\nu \rho} x_\rho \part_\nu \Phi^c,
~~~~\epsilon^\nu = \epsilon^{\nu \rho} x_{\rho}
\en
with $\epsilon^{\nu \rho} $ infinitesimal constant
Lorentz parameter. Substituting into (\ref{varSonshell})
yields
 \eq
  0=\de S= \int_M \epsilon^{\nu\rho} \dmu M^\mu_{~\nu\rho}  ~{\mbox{d}}^D x .
   \en
 \noi where
\begin{eqnarray}
M^\mu_{~\nu\rho} &=& {e \over 2} x_{[\nu } \part_{\rho ]}  \Phi \{\dmup \Phi\ds e^{-1}\}
+ {e \over 2} x_{[\nu } \part_{\rho ]}  \Phi_c \{ \dmup \Phi^c \ds e^{-1} \}\nonumber \\
&& - e \viel{a}{\mu} \Big( \Lcal_\star \star (e^{-1} x_{[\nu } \part_{\rho ]}  \Phi^a )\Big)  \nonumber \\
&& +  e \viel{a}{\mu} \Big[ T(\Delta) (X_c \Lcal_\star, \Xtilde^a (e^{-1} x_{[\nu }
\part_{\rho ]}  \Phi^c)) \nonumber \\
&& \quad\quad - T(\Delta) \Big( \part_{[\nu} \Phi, \unmezzo \Xtilde^a (\{\part_{\rho]} \Phi \ds e^{-1} \} )\Big) \nonumber \\
&&\quad\quad - T(\Delta) \Big( \part_{[\nu} \Phi^d, \unmezzo \Xtilde^a ( \{\part_{\rho]} \Phi_d \star e^{-1} \} )\Big) \nonumber \\
&&\quad\quad + S(\Delta) (\part_{[\nu} \Phi, \Xtilde^a (\part_{\rho ]}  \Phi \star e^{-1})) \nonumber \\
&& \quad\quad + S(\Delta) (\part_{[\nu} \Phi_d, \Xtilde^a (\part_{\rho ]}  \Phi^d \star
e^{-1})) .\label{Mmunurho}
\end{eqnarray}
\noi is the conserved angular momentum tensor.
In the commutative limit it reduces to the canonical
angular momentum tensor of the decoupled $\Phi$ and $\Phi^c$
fields.

 Again $\dmu M^\mu_{~\nu\rho}$ can be explicitly verified to vanish on shell.

\sect{Conclusions}

By means of an extension of the Moyal product, we have implemented
dynamical noncommutativity in $\Phi^{\star 4}$ theory (in fact all
the results hold also for $\Phi^{\star n}$), and simultaneously
restored global Lorentz symmetry. This we achieved by introducing
$x$-dependence in the definition of the $\star$-product in a
factorized way, cf. (\ref{theta}),  (\ref{twist}).

We have seen that the field equations of the resulting twisted
$\Phi^{*4}$ theory admit the vacuum solution  $\Phi = 0$,
$\viel{\mu}{a} \equiv \dmu \Phi^a = \de^a_\mu$. In particular when
$\viel{\mu}{a} = \de^a_\mu$ the $\star$-product between any two
functions reduces to the Moyal product.

We conclude with the following observations:
 \sk
 \noi {\bf 1)} The $\star$-product $f \star g$ differs from the ordinary product $fg$ by
 terms involving partial derivatives of $f$, $g$ and $\Theta^{\mu\nu}$ (containing $\viel{a}{\mu} $): then for slowly varying
 functions $f$, $g$ and $\viel{a}{\mu} $ the $\star$-product is well approximated by the ordinary product. To be more precise,
 if the functions involved are approximately constant in cells of linear dimension
 $ \sqrt{\theta}$ where $\theta$ is the order of magnitude of the $\theta^{ab}$
 dimensionful constant parameters, and if $\viel{a}{\mu}$ is of the order of
 unity, then the $\star$-product can be replaced by
 the ordinary (commutative) product. Indeed in this case the typical terms
 in $f \star g - fg$  satisfy
  \eq
  \theta^{ab} (\viel{a}{\mu} \dmu f)(\viel{b}{\nu} \dnu g) \approx
   (\sqrt{\theta} \part f)  (\sqrt{\theta} \part g) \approx 0
   \en

 \noi {\bf 2)} There are other ways to deform $\Phi^{\star 4}$ theory: indeed
  a kinetic term for the $\Phi$ field
  \eq
    \int [\viel{\mu}{a} \star X_a(\Phi) \star \viel{\nu}{b} \star X_b(\Phi)
    \star e^{-1} ] \eta^{\mu\nu}e\,{\rm d}^Dx
    \en
    \noi still reduces to the usual kinetic term in the
    commutative limit. Note that $\viel{\mu}{a} \star X_a(\Phi)$ is just the $\star$-Lie derivative along
    the vector field $\partial_\mu$ (see \cite{grav2}, Section 4). If we use this kinetic term in ${\cal
    L}_\star$, the resulting field equation for $\Phi$ becomes:

  \eq
  X_a ( \eta^{\mu\nu} \{ \viel{\mu}{b} \star X_b(\Phi) \ds e^{-1}\} \star
 \viel{\nu}{a}) +
   \frac{m^2}{2}\{ \Phi \ds e^{-1}\} +
    \frac{\lambda}{4!} \{\Phi\star\Phi \ds \{\Phi \ds e^{-1}\} \} =0 .
  \en
  \noi In this equation all products are $\star$-products. We have not found
   a similar improvement for $\Phi^c$,
  essentially because the rule (\ref{varstar}) involves ordinary
  products of the type $\de \Phi^c X_c f$. However ordinary
  products in the field equation (\ref{phicequation}) can be
  transformed into $\star$-products via the twist ${\cal F}$.
  Indeed, if we define

\eqa & & {\cal F} = e^{-\frac{i}{2}\theta^{ab}X_a \otimes X_b}
\equiv \rf^\alpha \otimes \rf_\alpha  \\
 & &  {\cal F}^{-1} = e^{\frac{i}{2}\theta^{ab}X_a \otimes X_b}
\equiv \overline{\rf}^\alpha \otimes \overline{\rf}_\alpha \ena

 \noi where $\rf^\al, \rf_\al, \overline{\rf}^\alpha , \overline{\rf}_\alpha$ are elements of the universal enveloping
 algebra of the $X_a$, then
  \eq
  g \star h =  \overline{\rf}^\alpha (g)  \overline{\rf}_\alpha (h)
   \en
   \noi so that
   \eq
   gh = \rf^\al (g) \star \rf_\al (h)
   \en

 \noi {\bf 3)}
The extension of our
results to include (noncommutative) gravity is under study. In this perspective
we notice that
the vector fields $X_a = \viel{a}{\mu} \dmu$ are invariant not
only under global Lorentz rotations, but also under general
coordinate transformations.

\sk

\sk \noi{\bf Acknowledgements} \sk

This work is partially supported by the Italian MUR under contract
PRIN-2005023102 {``Strings, D-branes and Gauge Theories''} and by
the European Commission FP6 Programme under contract
MRTN-CT-2004-005104 ``{Constituents, Fundamental Forces and
Symmetries of the Universe}''.

\sect{Appendix}

We collect here some formulas used to derive the results of
Sections 3 and 4.

\sk
 \noi {\bf Star product} \eqa & &
 f \star g \equiv fg + {i\over 2} \theta^{ab} (X_a f)(X_b g) + {1 \over 2!}
 \left( {i \over 2} \right)^2
\theta^{a_1 b_1} \theta^{a_2 b_2} (X_{a_1} X_{a_2} f)
 (X_{b_1} X_{b_2} g) + \cdots \nonumber \\ & &
  ~~~~~~~ \equiv e^\Delta (f,g) \label{starproduct}
  \ena
  \noi where powers of the bilinear operator $\Delta$ are defined
  as
  \eqa
  & & \Delta^n (f,g) \equiv \left( {i \over 2} \right)^n
\theta^{a_1 b_1} \cdots \theta^{a_n b_n} (X_{a_1} \cdots X_{a_n}
  f) (X_{b_1} \cdots X_{b_n} g) \\
  & & (\Delta^0 (f,g ) \equiv fg)
  \ena
  \noi {} From the definition (\ref{starproduct}) one finds the following identities
  (straightforward extensions of the identities
  derived in \cite{Pengpan} for the usual Moyal product):

\eqa f \star g &=& fg + X_a \Big[ {\exp( \Delta) -1 \over \Delta}
(f, \Xtilde^a(g)] \Big] ,\label{app1}\\ \lbrack f \ds g \rbrack
&\equiv & f \star g - g \star f = 2 X_a \Big[ {\sinh \Delta \over
\Delta}(f, \Xtilde^a g)\Big] , \label{app2}\\ \{ f \ds g \}
&\equiv& f \star g + g \star f = 2fg + 2 X_a \Big[ {\cosh \Delta
-1 \over \Delta} (f, \Xtilde^a g)\Big] , \label{app3} \ena
 \noi with $\Xtilde^a \equiv {i\over 2} \theta^{ab} X_b$.

\sk

 \noi {\bf Derivatives and variations} \eqa \de_{\Phi^c}\viel{a}{\mu}
&=& - \viel{a}{\nu} \viel{b}{\mu} \de_{\Phi^c}\viel{\nu}{b} =  -
\viel{a}{\nu} \viel{b}{\mu} \dnu \de \Phi^b = -  \viel{b}{\mu} X_a
(\de \Phi^b) ,\label{app4}\\ \dmu e &=& e \viel{a}{\nu} \dmu
\viel{\nu}{a} = e \viel{a}{\nu} \dnu \dmu \Phi^a = e X_a (\dmu
\Phi^a) , \label{app5}\\ \de_{\Phi^c} e &=&  e \viel{a}{\nu} \de
\dnu \Phi^a  = e \viel{a}{\nu} \dnu (\de \Phi^a) = e X_a (\de
\Phi^a) ,\label{app6}\\ \de_{\Phi^c} X_a &=& \de_{\Phi^c}
(\viel{a}{\mu} \dmu) = - \viel{b}{\mu} X_a (\de \Phi^b ) \dmu = -
X_a (\de \Phi^b) X_b \label{varX} ,\label{app7}\\ e X_a (f) &=&
\dmu ( e \viel{a}{\mu} f)  .\label{eX} \ena

\sk
In computing $\de_{\Phi^c}$ variations, the following identity is
useful:
\eq
\de_{\Phi^c} (f \star g) = - (\de \Phi^c X_c f) \star g - f
\star  (\de \Phi^c X_c g) + \de \Phi^c X_c ( f \star g) \label{varstar'}
\en
where the functions $f$ and $g$ do not depend on $\Phi^c$. This formula gives the $\de \Phi^c$ variation of a $\star$-product of two functions, due to the $\Phi^c$ fields contained in the definition of $\star$, and can be proved by considering the variations of the typical term in $f \star g$:
\eq
\de_{\Phi^c} [ (X_{a_1} \cdots X_{a_n} f)( \Xtilde^{a_1} \cdots
\Xtilde^{a_n} g)]
\en
By induction one proves easily that (\ref{varstar}) holds for
$\star$-products of an arbitrary number of factors:
\begin{eqnarray}
\de_{\Phi^c} (f \star g \star \cdots \star h) &=& - (\de \Phi^c X_c f) \star g \star \cdots \star h \nonumber\\
&& -f \star  (\de \Phi^c X_c g)\star \cdots \star h - f \star g \star \cdots \star  (\de \Phi^c X_c h) \nonumber \\
&& + \de \Phi^c X_c ( f \star g  \cdots \star h) . \label{varstarn}
\end{eqnarray}

%%%%%%%%%%%%
%%%%%%%%%%%%
\sk
 \noi {\bf Note:} a different method to compute $\de_{\Phi^c}$ variations
%and (\ref{varstar'})
is to recall that the $\Phi^a$ fields determine the invertible
transformation
$x\rightarrow \varphi(x)$, i.e., in coordinates
$x^\mu\rightarrow \Phi^a(x)$. Any expression
$V=V\left(\Phi, X_a(\Phi), X_aX_b(\Phi), ...\right)\vert_x$
that depends on the scalar field $\phi$ and its $X_a$
derivatives (like the potential term ${\phi^\star}^4$) can then be written as
\eq
V=V\left(\phi\circ\varphi^{-1}, (\phi\circ\varphi^{-1})_{a},
(\phi\circ\varphi^{-1})_{ab},...\right)\vert_{\varphi(x)}
\en
where the indices $a,b,...$ denote the partial derivatives
${\partial\over{\partial\Phi^a}}, ~{\partial\over{\partial\Phi^b}},...\,$.
Under the variation $\Phi^c\rightarrow \Phi'^c=\Phi^c+\de\Phi^c$ we have
\eqa
V&\rightarrow&
V\left(\phi\circ\varphi'^{-1}, (\phi\circ\varphi'^{-1})_{a},...\right)\vert_{\varphi'(x)}
\\&&=
V\left((\phi\circ\varphi'^{-1}\circ\varphi)\circ\varphi^{-1},
((\phi\circ\varphi'^{-1}\circ\varphi)\circ\varphi^{-1})_{a},...\right)\vert_{\varphi(\varphi^{-1}(\varphi'(x)))}~.\nonumber
\ena
In the last line the ${\Phi^c}$ variation has been rewritten as a  $\Phi$
variation and a coordinate transformation. Infinitesimally
these are given by
\eqa
\delta^{^{(\delta\varphi)}}_\Phi\Phi(x)&\equiv &(\Phi\circ\varphi'^{-1}\circ\varphi)(x)-\Phi(x)=-\delta\Phi^aX_a\Phi(x)~,\nonumber\\
\delta_x^{^{(\delta\varphi)}}x &\equiv &{\Phi^{-1}}^\mu(\varphi'(x))-x^\mu=\delta\Phi^aX_a(x^\mu)~.
\ena
We conclude that the $\delta_{\Phi^c}$ variation equals the
$\delta^{^(\delta\varphi)}_\Phi$ variation
 plus the variation generated by the vector field $\delta\Phi^aX_a$,
\eq
\delta_{\Phi^c}V=\delta^{^{(\delta\varphi)}}_\Phi V+\delta\Phi^aX_a V~.
\en

\vfill\eject

\end{document}